\documentclass[10pt,a4paper]{article}
\usepackage[a4paper]{geometry}
\usepackage[T1]{fontenc}
\usepackage{mathpazo}
\usepackage[latin1]{inputenc}							
\usepackage{amsmath}									
\usepackage{xcolor,graphicx}
\definecolor{bl}{rgb}{0.0,0.2,0.6}
\definecolor{nicered}{rgb}{.647,.129,.149}
\newcommand{\sgn}{\mathrm{sgn}}
\newcommand{\ud}{\mathrm{d}}
\newcommand{\Flip}[1]{\text{$\text{F}_{\text{#1}}$}}

\usepackage{hyperref}
\hypersetup{
    bookmarks=true,         
    pdftoolbar=true,        
    pdfmenubar=true,        
    pdffitwindow=false,     
    pdfstartview={FitH},    
    pdftitle={A statistical physics perspective on criticality in financial markets},    
    pdfauthor={Thomas Bury},     
    pdfnewwindow=true,      
    colorlinks=true,       
    linkcolor=bl,          
    citecolor=blue ,        
    filecolor=red!90,      
    urlcolor=blue ,          
}

\usepackage{sectsty}
\usepackage[compact]{titlesec}
\titleformat{\section}{\color{nicered}\large\bf}{\thesection}{1em}{}
\titleformat{\subsection}{\color{nicered}\normalsize\bf}{\thesubsection}{1em}{}
\titleformat{\subsubsection}{\color{nicered}\normalsize\bf}{\thesubsubsection}{1em}{}

\makeatletter							
\def\printtitle{
    {\color{bl} \flushleft \huge \@title\par}}		
\makeatother							

\title{A statistical physics perspective on criticality in financial markets}

\makeatletter							
\def\printauthor{
    {\hfill\parbox[b]{0.90\textwidth}{\flushleft \small \@author}}}				
\makeatother							

\author{%
	\textbf{\large Thomas Bury} \\[1\baselineskip]
    Service OPERA (CP194/5), Universit\'e libre de Bruxelles,\\
    Avenue F.D. Roosevelt 50, 1050 Brussels, Belgium\\
	Email:tbury@ulb.ac.be \\
	}

\usepackage{fancyhdr}
\usepackage{lastpage}	
\usepackage[runin]{abstract}		        
\abslabeldelim{:\quad}						%
\begin{document}
\printtitle

\printauthor

\begin{abstract}Stock markets are complex systems exhibiting collective phenomena and particular features such as synchronization, fluctuations distributed as power-laws, non-random structures and similarity to neural networks. Such specific properties suggest that markets operate at a very special point. Financial markets are believed to be critical by analogy to physical systems but few statistically based evidence have been given.
Through a data-based methodology and comparison to simulations inspired by statistical physics of complex systems, we show that the Dow Jones and indices sets are not rigorously critical. However, financial systems are closer to criticality in the crash neighbourhood.
\end{abstract}

\hrule
\footnotesize
\tableofcontents
\vspace{1em}
\hrule
\vspace{1em}
\normalsize
\section{Introduction}
\label{intro}

A hundred years ago, the Italian economist Pareto introduced the notion of a power-law describing wealth distribution. This is a major concept related to the notion of scale invariance which is widely used in finance and economics (fractional Brownian motion, detrended fluctuations analysis, volatility modelling, etc.). Scale invariance is crucial in finance because large absolute returns are power-law distributed \cite{Cont}. This lack of any characteristic scale is surprising at first glance but finds its foundation in the theory of complex systems. As complex systems composed of many correlated entities, financial markets exhibit collective behaviours like synchronization or non-random structure, propensity to self-arrange in large correlated structures as highlighted in \cite{Mant,OnnelaPRE,moi2}, large fluctuations \cite{Gabaix} and power-laws \cite{Cont}. Moreover it has been shown that financial networks share common properties with neural networks \cite{Petra}. One recovers those features in a class of models belonging to statistical physics, pairwise maximum entropy models which are particulary suited to capture collective behaviours. One knows that the market may exhibit some of the former features at a critical state, defined in a precise sense \cite{Huang} and that maximum entropy models may describe collective behaviours observed in neural networks \cite{Schn} and in financial markets \cite{moi2}. It is therefore tempting to think that financial markets are critical \cite{Sorn2} (in statistical physics sense) as it seems to be for neural networks \cite{fraiman,Egu,Tagl}.
It is not obvious how to validate empirically the presence of a critical state. Criticality was proposed for the approach of log-periodicity \cite{SorLog}. The phenomenological comparison to critical phenomena was done by substituting the temperature by the time which becomes therefore the control parameter \cite{Vande} but it is merely an analogy, log-periodicity should be understood as a dynamical feature rather than a second order phase transition. Indeed, several dynamical mechanisms generate log-periodicity \cite{SorDSI}. Criticality was also proposed for agent-based models exhibiting power-laws and volatility clustering at this particular state \cite{Lux,zhou,Moro,Yeung,Cha,Sav}. However different rules and models lead to the same qualitative stylized facts. There is still ambiguity since there are non-critical mechanisms which generate stylized facts \cite{Newman}. Furthermore, detecting the criticality is not the same task as modelling complex systems, even if relations obviously exist. A rigourous approach of criticality detecting is the \emph{inverse} (or data-based) approach.  A transition between scale dependence and scale invariance is highlighted \cite{Kiyono} in this way. Here, we also follow an inverse (starting from the data without initial assumptions) procedure described in \cite{Mora} and applied in \cite{Step}. This procedure is also inspired by statistical physics and provides several statistical tests of criticality.

We find that the considered financial systems are not strictly critical even if some signatures are observed. It is more likely that financial systems do not stay in the same regime and are closer to the criticality when the system gets closer to a crash. The critical scaling parameter (see hereafter) reaches its maximal value in the vicinity of the beginning of the crash. Namely, the response function to a shock (a shock can be a modification of exogenous variables or of the level of stochasticity) has a peak and its position scales with system size towards the operating point (at which the probability distribution is the empirical one) for European market places. The operating point of the Dow Jones is far from the critical one but the criticality could be reached if the size of the index is large enough. The distribution of rank of configurations is not a power-law if the system is well sampled and the entropy is not a linear function of the log-likelihood. Moreover, we use a pairwise maximum entropy model \cite{moi1,moi2} to check that the variance of the log-likelihood and the variance of the overlap parameter reach their maximum at x-axis coordinates in line with the empirical ones. We compare empirical results to simulations of a multivariate GARCH process and a Monte Carlo Markov Chain. They corroborate the empirical findings. Last, we give an interpretation of criticality in financial markets. These findings can be important in portfolio optimization which relies on the market structure (through the correlation matrix, for instance) and to figure out how market processes information which may eventually lead to a crash.

\section{Signatures of criticality}\label{sig}
A critical state can be thought as a state where the system lies at the threshold between order and disorder. If there is no uncertainty, markets are perfectly ordered and thus homogeneous (either positive or negative). In the opposite situation where uncertainty is maximal, markets are completely random and uncorrelated; the probability to observe a positive or negative return is equal to $1/2$ whatever the returns (positive or negative) of other market exchanges. A critical state is halfway these extreme states, letting markets on the edge of disorder and highly heterogenous.

Strictly, a critical state can be achieved only for infinite size systems. For finite systems, one will not observe divergences but we will still say \emph{critical} through the misuse of language and we should compare the empirical results to a finite version of a model which  may actually reach the criticality (the nearest neighbour Ising model in two dimensions for instance) as proposed in \cite{Step}.


Statistical physics provides several tests of criticality. The signatures detailed in \cite{Mora} will be briefly recalled. First, we define a \emph{financial system} as a set of stocks (or indices).  Relative stock returns are taken as random variable $r_{t}$ and can be rewritten as $r_{t}=\sgn(r_t)|r_{t}|$. Signs of stock returns are sometimes considered as uncorrelated and attract less attention. However correlations may appear in complicated (non-linear) fashion as synchronization during crises \cite{Jr}. It is interesting to study orientation (sign of returns) changes since Ising-like models are suited to describe collective behaviours. Moreover the nature of the relative return sign is more subtle than the one of simple independent random variables and can render the particular structure of financial markets \cite{moi1,moi2}. The net orientation is defined as $m(t)=N^{-1}\sum_{i}s_{i,t}$, if $m(t)>0$ the market trend is positive for the period $t$.
In order to study orientation changes, we consider a set of $N$ market indices or $N$ stocks described by binary variables $s_{i}\equiv \sgn(r_{i})$ ($s_{i}=\pm1$ for all $i=1,\cdots,N$). A system configuration will be described by a vector $\textbf{s}=(s_{1},\cdots,s_{N})$. The binary variable will be equal to one if the trend of the associated stock is positive and equal to $-1$ if not. A configuration $(s_{1},\cdots,s_{N})$ may also be thought as a binary version of the returns.

One can formally write the probability $P(\textbf{s})$ of finding the system in state $\textbf{s}$ as a Gibbs distribution $P(\mathbf{s})=\mathcal{Z}^{-1}e^{ \mathcal{U}(\mathbf{s})}$ and without loss of generality set $\mathcal{Z}$ to $1$ which leads to the definition of the utility function (or \emph{energy} $\mathcal{H}=-\mathcal{U}$, \emph{potential}, etc.) as the log-likelihood: $\mathcal{U}(\textbf{s})=\log P(\textbf{s})$. The rank $r(\mathrm{\textbf{s}})$ of a given configuration $\textbf{s}$ is defined as the number of configurations with a higher utility (more frequent) than the value associated to $\textbf{s}$.

\begin{figure}[ht!]
\begin{center}
\includegraphics[scale=1]{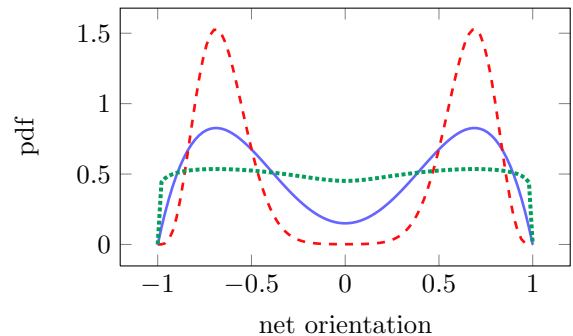}
\end{center}
\caption{Schematic illustration of the rescaling of a bimodal distribution as encountered in the Landau phenomenological theory of phase transition. The original probability density is illustrated by the full line at an arbitrary temperature $T^{*}$; the rescaled distributions are illustrated by the dashed line (at $T=0.25T^{*}$) and the dotted line (at $T=10T^{*}$).}
\label{fig:flat}
\end{figure}

A power-law $-\log P(\textbf{s})=\alpha \log r(\mathrm{\textbf{s}})+\mathrm{Cst}$ is a strong signature of criticality. In this framework, it is possible to obtain these quantities directly from a large enough sample and test the validity of Zipf's law. Another consequence of this law is the linearity of the Shannon entropy \cite{Step}, which measures the average \emph{surprise} or average \emph{log-likelihood}, expressed in term of an utility function \cite{Mora}. A weaker signature is the divergence of the variance of the likelihood at the operating point (in the limit of infinite number of entities). For finite systems, the variance of the likelihood should peak near the operating point if the system is in a critical state. This feature can also be checked directly from the data. The empirical relative frequencies are scaled as $P_{T}(\textbf{s})=P(\textbf{s})^{1/T}/\sum P(\textbf{s})^{1/T}$; the operating point corresponds to $T=1$. We note that for such a Gibbs distribution we have the identity

\begin{equation}\label{FD}
  R_{\mathcal{U}}=-\frac{\partial\langle \mathcal{U} \rangle}{\partial T}=T^{-2}\left[\langle \mathcal{U}^{2}\rangle-\langle \mathcal{U}\rangle^{2}\right]
=T\frac{\partial \mathcal{S}}{\partial T}
\end{equation}
where $\mathcal{S}(T)$ is the Shannon entropy $-\sum_{\{\mathbf{s}\}}P_{T}(\textbf{s})\log P_{T}(\textbf{s})$ of the rescaled distribution and the brackets stand for the average with respect to $P_{T}(\textbf{s})$.
In a statistical point of view, this extremum is the point where the deviation to equiprobability of events is the largest. Operating at this point involves that the variance of the log-likelihood reaches its largest value whereas for equiprobable events, the variance of the log-likelihood is equal to zero. A large variance of the log-likelihood also implies a large deviation from its mean value, the entropy, and thus large structural changes. The rescaling parameter $T$ can be thought as a randomness measure, changing this parameter leads to a reweighting of the empirical distribution. For $T>1$, the distribution will be flattened and closer to the uniform distribution as illustrated in Fig-\ref{fig:flat}. The entropy of the remaining distribution will thus be larger than the original one: the closer to the uniform distribution, the larger the entropy. We note that the expression $T\partial\mathcal{S}/\partial T$ is useful when direct sampling of probability distribution is feasible and the expression $T^{-2}\left[\langle \mathcal{U}^{2}\rangle-\langle \mathcal{U}\rangle^{2}\right]$ allows estimation through a Monte Carlo simulation even if direct sampling is unfeasible. When direct sampling is feasible, one can estimate the empirical distribution as $P(\mathbf{s})=M^{-1}\sum_{i=1}^{M}\delta_{\mathbf{s}_{i},\mathbf{s}}$ where $M$ is the sample length, compute the scaled distribution $P_{T}(\textbf{s})$ for any value of $T$ and then use the relation  $T\partial\mathcal{S}/\partial T$ for the empirical derivation of the response function.

\section{Sampling indices and stock exchanges}
\begin{figure}[ht!]
\begin{center}
\includegraphics[scale=1]{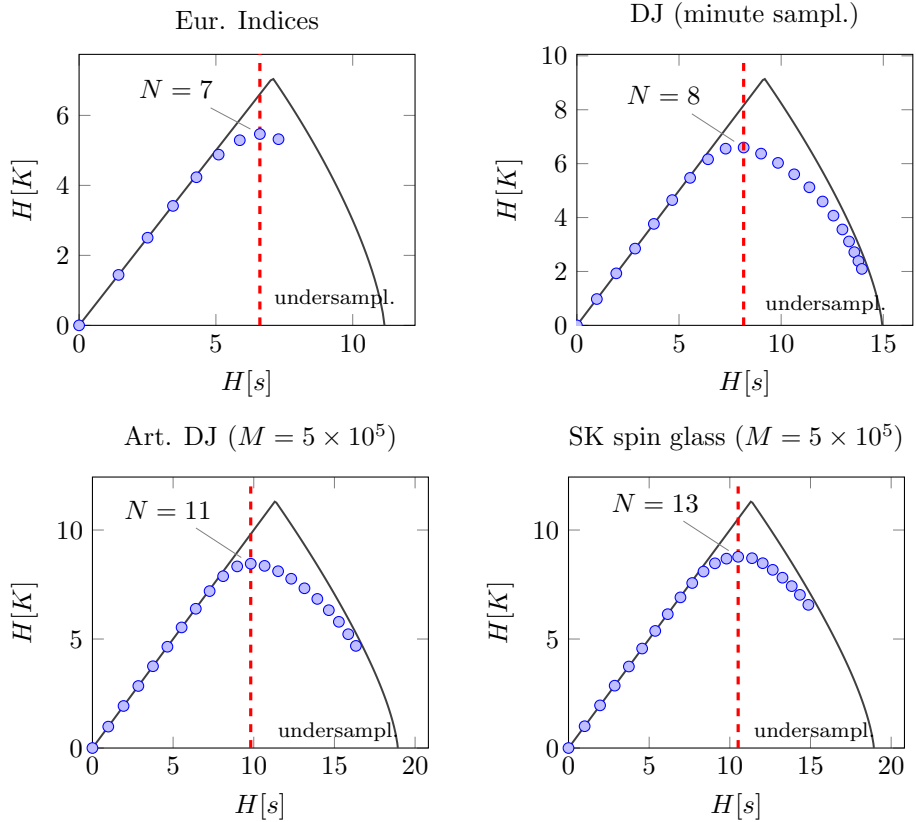}
\end{center}
\caption{Statistical significance of data sets. The configurations distribution $P[\mathbf{s}]$ is correctly sampled in the left part of the plane $\{H[K],H[\mathbf{s}]\}$, delimited by the dashed line. The full line stands for theoretical relation, $H[K]$ as a function of $H[\mathbf{s}]$. The dots stand for empirical values for each data set, as the system size increases (from left to right in the plane $\{H[K],H[\mathbf{s}]\}$). The right-bottom panel illustrates the results for a SK spin glass of size $N=25$ near the criticality. For the European indices set, each point is calculated by averaging over ${8 \choose N}$. For the Dow Jones (daily and minute samplings), each point is calculated by averaging over 100 sets of $N$ randomly chosen stocks.}
\label{fig:StatSig}
\end{figure}

We observed opening and closing prices of 8 European indices (AEX, BEL, CAC, DAX, EUROSTOXX, FTSE, IBEX, MIB), the sample length $M$ is 2300 trading days (approximatively nine trading years englobing two global crises, 2002-2011 period). We consider European stock exchanges because some issues (debt crisis, etc.) are specific to these market places and to ensure the simultaneity of time series. We also observed the stocks of the Dow Jones index during $3\times 10^{4}$ trading minutes and at daily sampling from 2002 to 2011. We consider two different time-scales to explore the differences when the correlations decrease. According to the Epps effect \cite{Epps}, we expect that systems sampled at low frequency (daily sampling) should be closer to the criticality than the systems sampled at larger frequencies (minute sampling, for instance). Positive returns are set to $1$ and negative returns to $-1$. The first sample is ten times larger than the number of possible configurations. Indeed, there are two possible values for each variable $s_{i}$, thus they are $2^{N=8}=256$ configurations. The second sample is not large enough for a satisfactory probability estimation (and thus a direct estimation of entropy).
Since entities may be strongly correlated, it is not obvious to know if the configurations are well sampled or not. In case of strongly correlated entities, the relevant region in the configurations space is narrow in comparison to independent entities. If the true configurations distribution is sharply peaked, there are only few relevant states. In this situation, a small ($M<2^N$) sample is enough to sample properly the configurations distribution. In the opposite case where entities are independent, every configuration has the same statistical weight and the sample size must be large ($M\gg 2^N$).
It is crucial to identify the maximum number entities one should consider to avoid undersampling of the configurations distribution $P[\mathbf{s}]$ because power-laws occur spontaneously in the undersampling regime \cite{MMR}. In particular, Zipf's law is only a genuine feature if $P[\mathbf{s}]$ is well sampled. To assess the maximum number of entities to consider in the analysis, we follow the procedure described in \cite{MMR}.
The limit between proper sampling and undersampling is defined by the coordinates of the maximum of $H[K]$ in the plane $\{H[K],H[\mathbf{s}]\}$ where $H[\mathbf{s}]$ is the entropy of the empirical configurations frequencies and $H[K]$ is the entropy of the random variable $K_{i}=K(\mathbf{s}_{i})$ which is the number of times the configuration $\mathbf{s}_{i}$ is observed in the sample. Beyond this point, $H[K]$ decreases when $H[\mathbf{s}]$ increases which means that configurations are sampled (approximatively) the same number of times.
Briefly, given a sample of $M$ independent configurations $(\mathbf{s}_{1},\cdots,\mathbf{s}_{M})$, the empirical distribution of the configurations is $\hat{p}_{\mathbf{s}}\equiv P[\mathbf{s}_{i}=\mathbf{s}]=M^{-1}\sum_{i=1}^{M}\delta_{\mathbf{s}_{i},\mathbf{s}}$. The distribution of the random variable $K_{i}$, corresponding to the number of times the configuration $\mathbf{s}_{i}$ occurs in the sample, is written $P[K_{i}=k]=k\,m_{k}/M$ where $m_{k}=\sum_{\{\mathbf{s}\}}\delta_{k,M\hat{p}_{\mathbf{s}}}$ is the number of configurations that are sampled exactly $k$ times in the sample. Their entropies are

\begin{eqnarray}
  H[\mathbf{s}]&=& -\sum_{\{\mathbf{s}\}}\hat{p}_{\mathbf{s}}\log \hat{p}_{\mathbf{s}}=-\sum_{k}\frac{k\,m_{k}}{M}\log\frac{k}{M}\\
  H[K]&=& -\sum_{k}\frac{km_{k}}{M}\log\frac{k\,m_{k}}{M}=H[\mathbf{s}]-\sum_{k}\frac{k\,m_{k}}{M}\log m_{k}
\end{eqnarray}
These quantities can be evaluated to obtain the statistical significance of each data set. The points in Fig-\ref{fig:StatSig} have been obtained by considering increasing system size. Each point is obtained by this mean and by averaging over several sets of randomly chosen entities (see the caption). Moreover, the theoretical limit is given by the most informative samples (full lines in Fig-\ref{fig:StatSig}) which are those maximizing $H[K]$ with respect to $\{m_{k},k>0\}$ and satisfying the constraints $H[\mathbf{s}]\leq N$, $\sum_{k}k\,m_{k}=M$ and $H[K]\leq H[\mathbf{s}]$ since the random variable $K$ is a function of $\mathbf{s}$ (see \cite{MMR} for the complete discussion and derivation).

The statistical significance is illustrated in Fig-\ref{fig:StatSig} for each data set and for artificial data simulated by fitting a pairwise maximum entropy model (see \ref{sec:MCMC}). We simulate also a time series of a Sherrington-Kirkpatrick (SK) spin glass of size $N=25$ near the criticality. The European indices set is correctly sampled up to 7 indices, the Dow Jones at minute up to 8 stocks. Increasing 15 times the sample length $M$, allows to consider up to $N=11$ entities.

A qualitative observation is that if entities are highly correlated (low stochasticity), almost all observed configurations (words) should be such that the mean orientation $m(t)=N^{-1}\sum_{i}s_{i,t}$ is non zero. One expects a $H[\mathbf{s}]$ significantly lower than the theoretical upper bound $N$ and $H[K]\simeq H[\mathbf{s}]$ since few different configurations are observed. On the other hand, nearly independent entities do not favour any value of $m(t)$. The configuration distribution $P[\mathbf{s}]$ should be approximatively uniform, $H[\mathbf{s}]$ should be close to $\mathrm{min}(N,\log_{2} M)$ for large system sizes and $H[K]$ should be small since each configuration is observed approximatively a same number of times.
From pairwise maximum entropy (maxent) models \cite{Fisher}, one knows that criticality is a regime where no net orientation is observed but where fluctuations are the largest. We expect that the sampling of a truly critical regime should return a situation halfway between the two previous extreme cases, as illustrated in Fig-\ref{fig:StatSig} for the SK-model.

\begin{figure}[ht!]
\begin{center}
\includegraphics[scale=0.9]{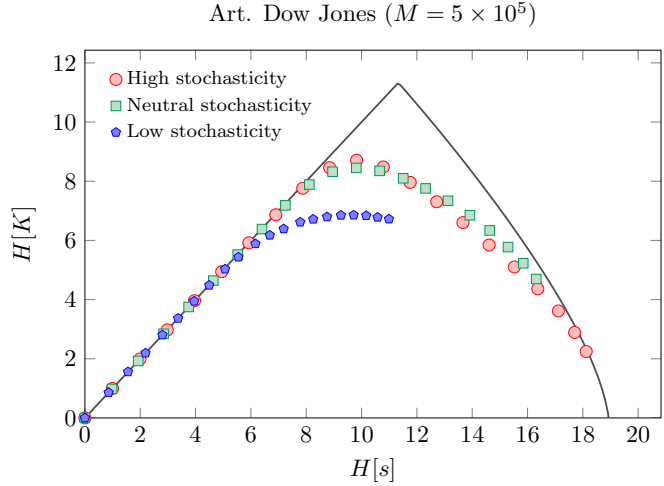}
\end{center}
\caption{Statistical significance of the Dow Jones data set for different levels of stochasticity. The full line stands for theoretical relation, $H[K]$ as a function of $H[\mathbf{s}]$. The dots stand for artificial data generated with a pairwise maxent model fit on the Dow Jones data with 1 third larger stochasticity than actual one. The squares illustrate artificial data with the same level of stochasticity than the actual one and the pentagons illustrate data generated with 1 third lower stochasticity.}
\label{fig:StatSigDJ}
\end{figure}

After fitting a pairwise maxent model, we record artificial data for the Dow Jones varying the stochasticity by 1 third smaller and larger than the actual one. The results are illustrated in Fig-\ref{fig:StatSigDJ}.
It seems that the Dow Jones (minute sampling) is rather disordered, we will check this in detail hereafter.

\section{Results}
\begin{figure}[ht!]
\begin{center}
\includegraphics[width=\textwidth]{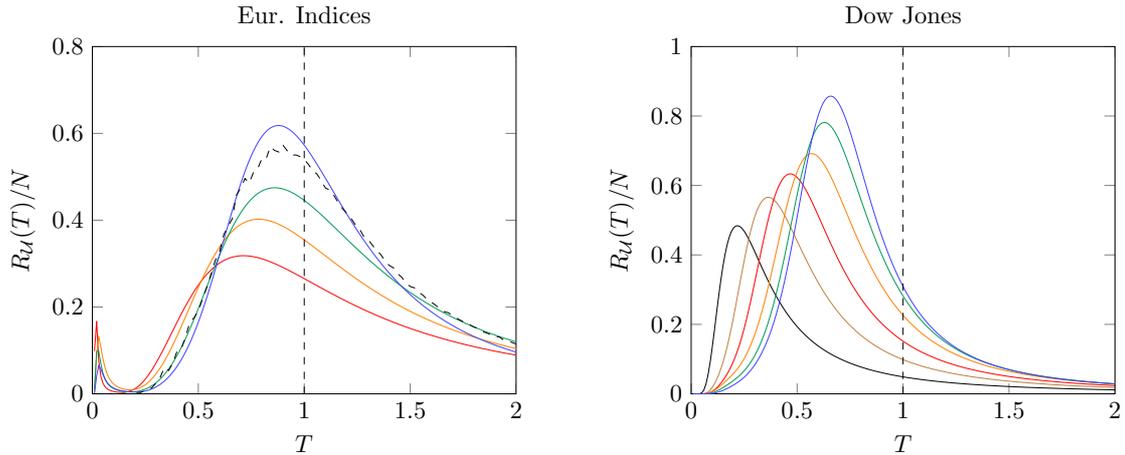}
\end{center}
\caption{Variance of the log-likelihood for the European indices set (left) and for the Dow Jones at minute sampling (right) vs the rescaling parameter. The peak moves from left to right when we consider larger sets. For the European set, we plot the variance for sizes $N=2,\,4,\,5,\,8$. The dashed curve is a Monte Carlo simulation (see \ref{sec:MCMC}) for $N=8$. For the Dow Jones, we consider $N=2,\,4,\,6,\,8,\,10,\,12$, the last two values are not statistically significant. These curves have been obtained by direct sampling of the probability (and entropy) and by using the relation $T\partial\mathcal{S}/\partial T$, see section \ref{sig} for details.}
\label{fig:hc}
\end{figure}

In the following, we check if the signatures of criticality are observed in the considered data sets. The variance of the log-likelihood is illustrated in Fig-\ref{fig:hc}.


\begin{figure}[ht!]
\begin{center}
\includegraphics[width=\textwidth]{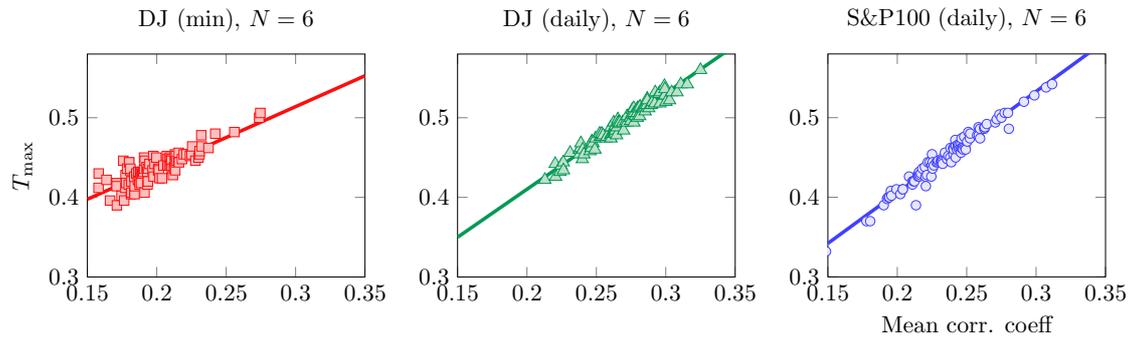}
\end{center}
\caption{The critical scaling parameter (x-axis coordinate of the maximum of the response function $R_{\mathcal{U}}(T)$) versus the mean correlation coefficient of the considered set of $N=6$ randomly chosen entities. The results are illustrated for the Dow Jones at minute (squares, left panel) and daily samplings (triangles, center panel) and for the S$\&$P100 (circles, right panel). The size $N=6$ is chosen consistently with the latter analysis of statistical significance.}
\label{fig:TcriCorr}
\end{figure}

We can observe that the peak position scales with the system size, moving from left to right towards the operating point $T=1$ and that the maximum value of the variance becomes larger when the number of entities increases. For a given and fixed size, one expects a larger value of the critical scaling parameter for sets (of $N$ randomly chosen entities) with a larger mean correlation coefficient \cite{Fisher,moi1}. We consider 100 sets of $N=6$ randomly chosen entities for the Dow Jones (daily and minute samplings) and for the S$\&$P100. The results illustrated in Fig-\ref{fig:TcriCorr} suggest a roughly linear relation between the critical scaling parameter $T_{\mathrm{max}}$ and the mean correlation coefficient. Any further results will thus be averaged over several sets for each considered size.

\begin{figure}[ht!]
\begin{center}
\includegraphics[scale=1]{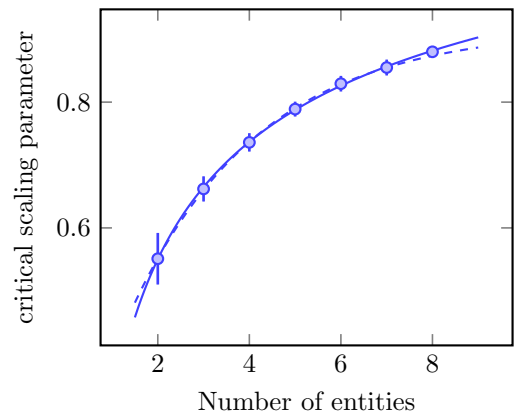}
\end{center}
\caption{Value of the scaling parameter at which response function $R_{\mathcal{U}}$ reaches its maximum value vs the number of entities $N$. Mean values $\bar{T}$ and error bars (1 standard deviation on $\bar{T}$) are computed over $\binom{8}{N}$ samples for European indices set. The full line stands for a power fit and the dashed line stands for an exponential fit on the first seven values. The response function is calculated using the relation (\ref{FD}).}
\label{fig:Tcri}
\end{figure}

\begin{figure}[ht!]
\begin{center}
\includegraphics[scale=1]{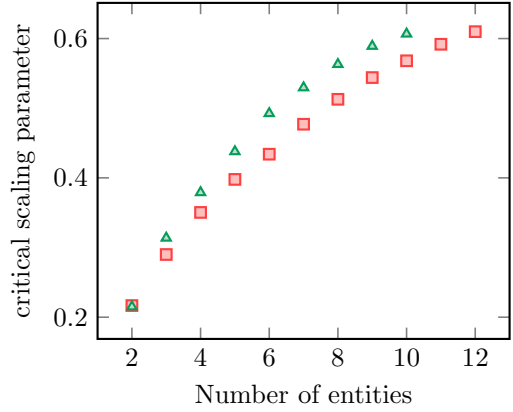}
\end{center}
\caption{Value of the scaling parameter at which response function $R_{\mathcal{U}}$ reaches its maximum value vs the number of entities $N$. Mean values are computed over 100 sets of $N$ randomly chosen stocks for the Dow Jones at daily (triangles) and minute (squares) samplings.}
\label{fig:Tcri2}
\end{figure}

To formalize the relation $T_{\mathrm{max}}=T_{\mathrm{max}}(N)$, we compute the value of the scaling parameter at which response function $R_{\mathcal{U}}$ reaches its maximum value for different sets of $N$ randomly chosen entities. Results are illustrated in Fig-\ref{fig:Tcri} for the European indices set and in Fig-\ref{fig:Tcri2} for the Dow Jones (daily and minute samplings).

The power and exponential fits return an asymptotic critical scaling parameter respectively equal to $1.38$ and $0.92$.

An exponential fit, on size up to $N=8$, of the DJ (min) returns an asymptotical critical parameter equal to $0.70$ and equal to $0.74$ if we fit up to $N=12$ (but the latter value is not trustful since the system is undersampled for $N>8$).
An exponential fit, on size up to $N=6$,  of the DJ (daily) returns an asymptotical critical parameter equal to $0.71$ and equal to $0.72$ if we fit up to $N=10$, (but the latter value is not trustful since the system is undersampled for $N>6$). Furthermore, even in the undersampled regime, we observe an increase of the critical scaling parameter.

\begin{figure}[ht!]
\begin{center}
\includegraphics[scale=1]{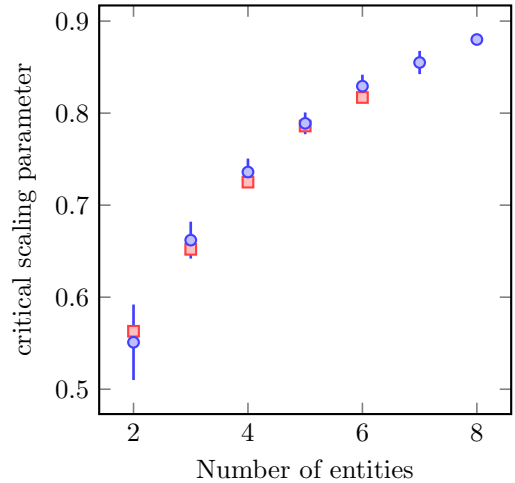}
\end{center}
\caption{Value of the scaling parameter at which the response function $R_{\mathcal{U}}$ reaches its maximum value vs the number of entities $N$. Mean values and error bars (1 standard deviation) are computed over $\binom{8}{N}$ samples for European indices set (circles). The squares illustrate results for the same sets with scaled sample length $L(N)=2^{N+3}$ and averaged over 5 different time-windows.}
\label{fig:TcriScale}
\end{figure}

Larger correlations measured when size ($N$) increases may be a spurious effect due to the consideration of a particular time interval. One can perform the same study by changing size and scaling sample length simultaneously and considering different time-windows. For the set of European indices, we chose sample length $L(N)=2^{N+3}$ such that  $L(8)\simeq L_{\mathrm{max}}=2300$ and we average the results on 5 different time-windows. Results are illustrated in Fig-\ref{fig:TcriScale}. Each point (square) falls into the confidence interval of the constant size results excepted the last one ($N=6$). Larger correlations for increasing size is thus a genuine feature.

\begin{figure}[ht!]
\begin{center}
\includegraphics[scale=1]{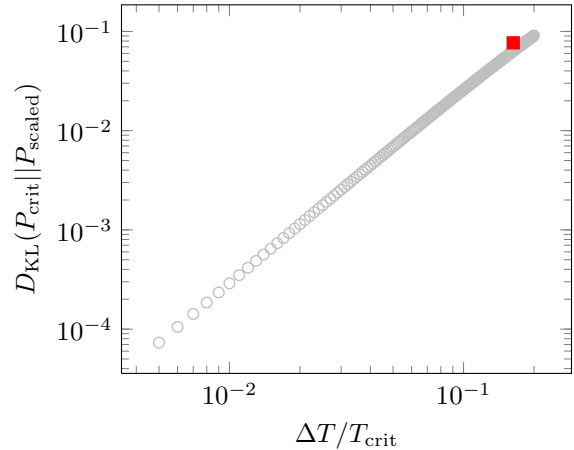}
\end{center}
\caption{Kullback-Leibler divergence between the critical and the scaled distributions for the two dimensional square lattice nearest-neighbours Ising model $N=9$ (light grey circles) and for the set of 8 European indices (square).}
\label{fig:DKL}
\end{figure}

As no inference method have been used, we expect that the Kullback-Leibler divergence (KLD) $D_{\mathrm{KL}}(P_\mathrm{crit}||P_\mathrm{emp})$ between the critical distribution $P_{T=T_{\mathrm{max}}}(\mathbf{s})$ (such that the maximum value of $R_{\mathcal{U}}$ is reached at $T_{\mathrm{max}}$) and the empirical distribution $P_\mathrm{emp}(\mathbf{s})$ should be of the same order of magnitude than for a truly critical system operating at $T_{\mathrm{crit}}+\Delta T$. The relative deviation $\Delta T/T_{\mathrm{crit}}$ and $(T_{\mathrm{op}}-T_{\mathrm{max}})/T_{\mathrm{max}}$ being equal (by definition $T_{\mathrm{op}}=1$). Following \cite{Step}, a reasonable benchmark is the two dimensional square lattice nearest-neighbours Ising model with periodic boundaries of size $N=9$. The response function $R_{\mathcal{U}}$ reaches its maximum value at $T_{\mathrm{crit}}=2.40$. We compute the exact distribution $P_\mathrm{crit}$ and the KLD with the scaled distribution $P_\mathrm{scaled}=P_\mathrm{crit}^{1/(1+x)}$ where $x=(T_{\mathrm{crit}}-T)/T_{\mathrm{crit}}$.
We found $T_{\mathrm{max}}=0.88$ and $D_{\mathrm{KL}}(P_\mathrm{crit}||P_\mathrm{emp})=0.070$ for empirical data (European indices). The results for the Ising model are illustrated in Fig-\ref{fig:DKL}.

For both systems, the results are similar. Furthermore, we simulated (see \ref{sec:MCMC}) artificial binary returns with a Monte Carlo Markov Chain ($1\times 10^{4}$ equilibrations steps and $2.3\times 10^3$ recorded configurations for $N=8$) using a pairwise maximum entropy model fitted on the data. We obtained an absolute net orientation $\hat{|m|}=0.812\pm0.010$ (1 standard deviation). The empirical value is $\langle|m|\rangle=0.726$, not included in the confidence interval but near a critical state, a slight change in inferred parameters may leads to significant change of observables estimated by simulations \cite{Mastro}. To quantify the effect of a small reconstruction error on the estimated observable, we inferred Lagrange parameters with a regularized pseudo-maximum likelihood (see \ref{sec:rPML} for details) and we shifted slightly the parameters such that $\Delta=0.015$, consistently with \cite{Aurell}. The reconstruction error is $\Delta=\sqrt{N}\langle (J_{ij}-J_{ij}^{\mathrm{true}})^{2}\rangle^{1/2}$ and quantifies the ratio between the root mean square error of the reconstruction and a canonical standard deviation. We obtained $10\%$ of relative deviation between the two estimations of $|m|$. The empirical and critical values of $|m|$ are thus similar.

\begin{figure}[ht!]
\begin{center}
\includegraphics[scale=1]{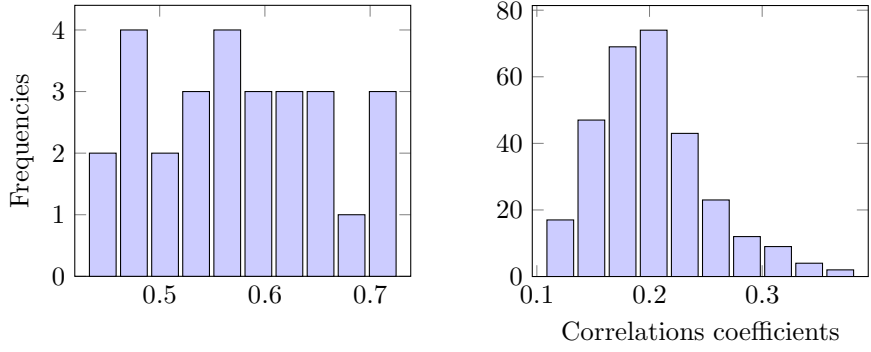}
\end{center}
\caption{Frequencies of correlation coefficients between European stock exchanges (left) and stocks of the Dow Jones index (right).}
\label{fig:CorrCoef}
\end{figure}

The European market places seem to operate near the point corresponding to the maximum of the variance of the log-likelihood while for the Dow Jones, the critical scaling parameter seems to be far away from the operating point $T_{\mathrm{op}}=1$ in the range of considered sizes. In Fig-\ref{fig:TcriArt}, we extend this plot for larger sizes by simulating artificial data (see hereafter). This may be explained by larger correlation coefficients between stock exchanges than between stocks of the Dow Jones as illustrated in Fig-\ref{fig:CorrCoef} and by the Epps effect (decreasing correlation magnitude with decreasing time-scale)\cite{Epps}.

Another observation is that the so-called critical exponent of the variance is equal to zero for each curve illustrated in the left panel of Fig-\ref{fig:hc} in agreement with the mean-field value of the Ising model at the critical temperature. The critical exponent can be obtained by taking the limit $\lim_{\epsilon\rightarrow0^{+}}\log R_{\mathcal{U}}(\epsilon)/\log \epsilon$ where $\epsilon=(T-T_{\mathrm{max}})/T_{\mathrm{max}}$ and $T_{\mathrm{max}}$ is such that $R_{\mathcal{U}}(T)$ reaches its maximum at this point \cite{Stanley}.

\begin{figure}[ht!]
\begin{center}
\includegraphics[width=\textwidth]{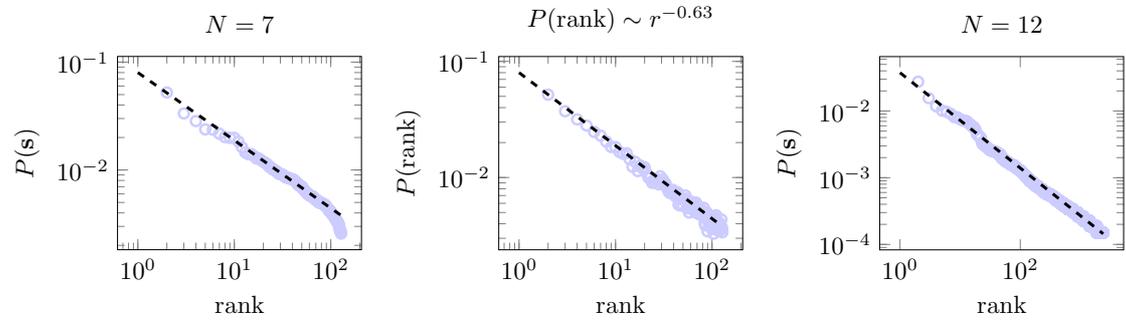}
\end{center}
\caption{From left to right: empirical relative frequencies of configurations vs the configurations rank of the observed time series for a set of 7 randomly chosen stocks of the Dow Jones (min), artificial rank distribution for a real power-law and empirical relative frequencies for a set of 12 randomly chosen stocks of the Dow Jones. The fit (dashed line) is obtained with the maximum likelihood estimator.}
\label{fig:zipf}
\end{figure}

We also study the distribution of the configuration rank. In order to know if we should reject or not Zipf's law, we perform a modified version (discrete power-law with a natural upper bound due to the finite number of configurations) of the statistical test described in \cite{Clauset}, see \ref{sec:DPL} for details. If the p-value is smaller than $0.05$, the power-law hypothesis is ruled out and for p-value close to one, we can consider it as a good distribution \emph{candidate} (without guarantee that it is the \emph{correct} distribution). The empirical rank distribution is illustrated in Fig-\ref{fig:zipf}.

Test results for different sets are reported in Table-\ref{tab:KS}. The considered size should not exceed $N=8$ for empirical data to have a good estimate of the distribution $P[\mathbf{s}]$ by direct sampling. As expected, the power-law test outcomes depend on the system size. For the Dow Jones, the power-law is rejected when the system is properly sampled whereas in the undersampling regime the power-law is not rejected. As detailed in \cite{MMR}, the power-law is the most informative distribution when the distribution $P[\mathbf{s}]$ is undersampled.

\begin{table}[!ht]
\caption{Statistical test of power-law hypothesis for sets of $N$ randomly chosen stocks of the Dow Jones ($3\times10^4$ points at minute sampling). We reported the maximum likelihood estimator $\hat{\alpha}$ of the power-law exponent $\alpha$ and its standard deviation $\sigma_{\alpha}$, the Kolmogorov-Smirnov statistic (D) and the p-value. One does not reject the power-law hypothesis if the p-value is larger than 0.05.}
\label{tab:KS}
\begin{center}
\begin{tabular}{ccccc}
  \hline
  $\#$ of stocks & $\hat{\alpha}$ & $\sigma_{\alpha}$ & D      & p-val \\\hline
  6                  &                &                   & 0.0119 & 0.00 \\
  7                  &                &                   & 0.0117 & 0.00 \\
  8                  &                &                   & 0.0194 & 0.00 \\
  9                  & 0.6654         & 0.0038            & 0.0147 & 0.10 \\
  10                 & 0.6584         & 0.0035            & 0.0164 & 0.36 \\
  11                 & 0.7192         & 0.0027            & 0.0210 & 0.87 \\
  12                 & 0.7441         & 0.0025            & 0.0292 & 0.96 \\
  13                 & 0.7699         & 0.0024            & 0.0290 & 0.98 \\
  \hline
\end{tabular}
\end{center}
\end{table}


The maximum likelihood estimator (MLE) of the exponent is derived by the maximization of the log-likelihood

\begin{equation}
\ln L(\alpha)= -\alpha \sum_{i=1}^{N}\ln x_{i}-N \ln\left(\sum_{x=1}^{x_{\mathrm{max}}}x^{-\alpha}\right)
\end{equation}

where $x_{\mathrm{max}}$ is the upper bound. The standard deviation of this MLE is obtained by taking the expansion of the likelihood up to second order (Gaussian approximation). It reads

\begin{equation}
  \sigma_{\alpha_{\mathrm{MLE}}}=\frac{1}{\sqrt{N\left[\frac{\zeta^{''}(x_{\mathrm{max}},\alpha_{\mathrm{MLE}})}
  {\zeta(x_{\mathrm{max}},\alpha_{\mathrm{MLE}})}-\left(\frac{\zeta^{'}(x_{\mathrm{max}},\alpha_{\mathrm{MLE}})}
  {\zeta(x_{\mathrm{max}},\alpha_{\mathrm{MLE}})}\right)^{2}\right]}}
\end{equation}

where $\zeta(x_{\mathrm{max}},\alpha)=\sum_{x=1}^{x_{\mathrm{max}}}x^{-\alpha}$ and the prime stands for the derivative with respect to $\alpha$.

\begin{figure}[ht!]
\begin{center}
\includegraphics[scale=1]{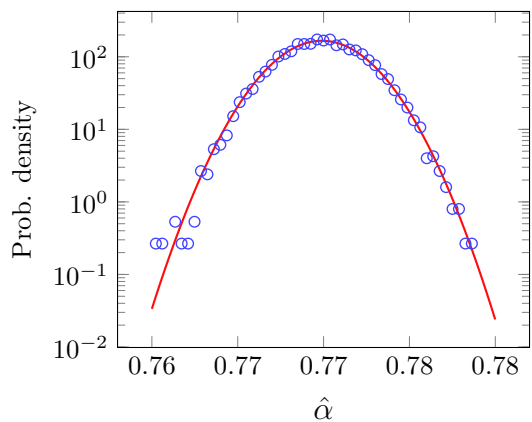}
\end{center}
\caption{Empirical pdf of the MLE estimator for the size $N=13$ and $10^{4}$ tests (circles). The Gaussian approximation is illustrated by the full line.}
\label{fig:BetaSD}
\end{figure}

The empirical probability density function (pdf) of this estimator for  $N=13$ and $10^{4}$ tests and its Gaussian approximation are illustrated in Fig-\ref{fig:BetaSD}.

As a complement to the later analyses, we study the linearity of the entropy expressed as a function of the utility.
Zipf's law induces a linear relation between entropy and the log-likelihood \cite{Mora}. The strict linearity can be achieved at a single value of the utility (as for the 2D nearest neighbor Ising model) or for \emph{any} value of the entropy if the distribution of the rank is a power-law \cite{Mora}. The expansion of the entropy around the mean utility $U$ is written (where $U$ is the notation for $\langle\mathcal{U}\rangle$)

\begin{equation}\label{LinEntr}
  S(\mathcal{U})\simeq S(U)-\frac{1}{T}(\mathcal{U}-U)+\frac{1}{2T^{2}R_{\mathcal{U}}} (\mathcal{U}-U)^{2}
\end{equation}

For ranks distributed following a power-law, the quadratic and higher order terms are sub-intensive; the entropy should be a linear function of the utility \cite{Step}.

%

\begin{figure}[ht!]
\begin{center}
\includegraphics[width=\textwidth]{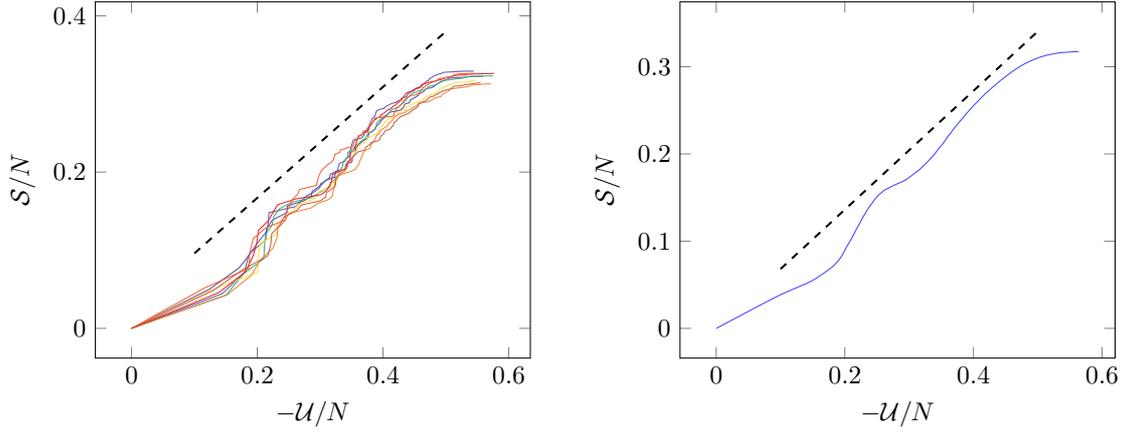}
\end{center}
\caption{Left: Shannon entropy vs the opposite of the log-likelihood for several sets of 7 stocks (randomly chosen) of the Dow Jones index (min). Right: the average entropy-utility relation $\mathcal{S}(-\mathcal{U})$ for 100 sets of 7 randomly chosen stocks. The dashed line is the best linear fit with slope equal to 0.71 and 0.68 respectively.}
\label{fig:EntrLin}
\end{figure}

We check this property for several sets of 7 randomly chosen stocks of the Dow Jones Index. We compute the average entropy-utility relation $\mathcal{S}(-\mathcal{U})$ for 100 sets of 7 randomly chosen stocks, the results are illustrated in Fig-\ref{fig:EntrLin}.

\begin{figure}[!ht]
\begin{center}
\includegraphics[scale=1]{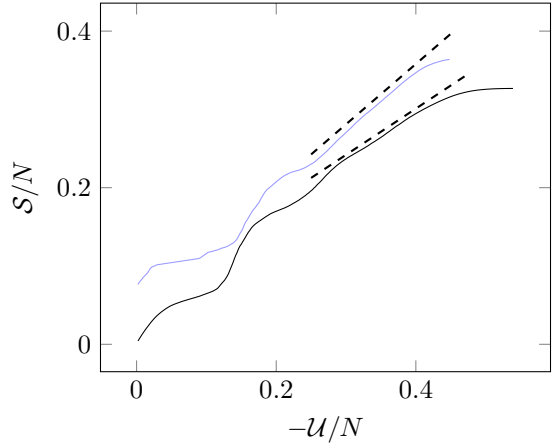}
\end{center}
\caption{Shannon entropy vs minus log-likelihood for 100 sets of 9 randomly chosen stocks. Artificial returns are simulated with a multivariate GARCH(2,2) process (light line) and with a pairwise maxent model (bold line). The dashed lines are a linear fit on a restricted range.}
\label{fig:EntrLinGARCH}
\end{figure}

We measured the relative non-linearity \cite{emancipator}, the typical value is $0.053$ (equal to zero if the function is exactly linear). The typical value of the slope is $0.71$. We also simulate $5\times 10^{5}$ artificial returns with a multivariate GARCH(2,2) and pairwise maxent processes fitted on the data. The entropy dependence on the log-likelihood is illustrated in Fig-\ref{fig:EntrLinGARCH}. The relative non-linearity is $0.032$ and $0.035$, the slope is equal to $0.77$ and $0.59$ respectively. For larger sample size the entropy is not linear either, however in a restricted utility range ($[0.3,0.4]$, about $10\%$ of the possible values of the utility, for instance) the entropy is almost linear (as measured by the relative non linearity).

As suggested by the Zip's law checking, the entropy is not a linear function of the log-likelihood. However, we can not reject the possibility of linearity in a restricted range or zero curvature in a single point as for the 2D nearest neighbour Ising model.

\begin{figure}[!ht]
\begin{center}
\includegraphics[scale=1]{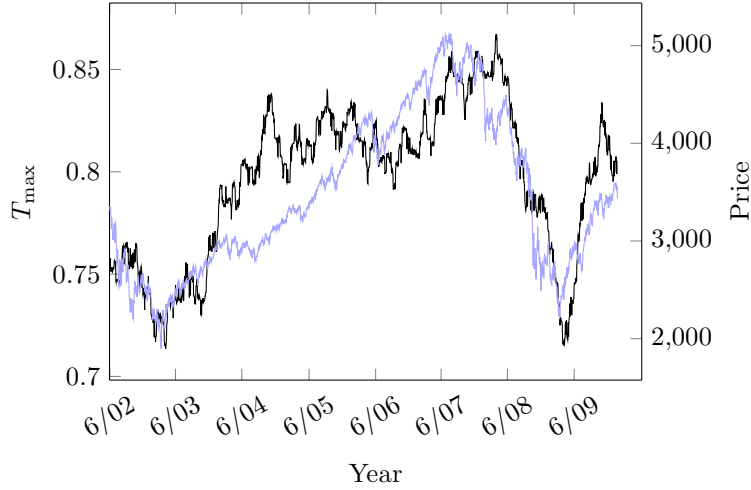}
\end{center}
\caption{Critical rescaling parameter $T_{\mathrm{max}}$ for 6 European indices (black curve, left ordinate) and the normalized sum of indices (light blue curve, right ordinate). The critical rescaling parameter is empirically estimated on a sliding window of $2^{N+2}$ trading days translated by $1$ trading day each step.}
\label{fig:TcritTime}
\end{figure}

Last, as the returns are believed to be non-stationary with volatility clustering (often modeled by a GARCH process), we study the evolution of the critical rescaling parameter $T_{\mathrm{max}}$ (at which the variance of the log-likelihood reaches its maximum value). As expected, for fixed size, $T_{\mathrm{max}}$ increases during the growth period and reaches its maximal value in vicinity of the crash beginning (when fluctuations are the largest) as illustrated in Fig-\ref{fig:TcritTime} and gets closer to $T_{\mathrm{op}}$.

\section{Link to maximum entropy models}\label{sec:MEM}
In the following, we use an inference procedure to check if the existence of a critical state is supported. One can show that the pairwise maximum entropy model (maxent) is a consistent statistical model when the aim is to study collective behaviours rather than to give a precise model of the market \cite{moi1,moi2}. Rather than making specific assumptions of the underlying dynamics, we build a model which is consistent with the recorded data and the observed structure. This maxent model is directly linked to the former discussion since spin glasses and neural networks are also represented by pairwise maxent models which actually exhibit critical states. In this framework the configurations distribution $P[\mathbf{s}]$ is rewritten as a Gibbs distribution

\begin{equation}
p_{2}(\textbf{s})=\mathcal{Z}^{-1}\exp\left(\frac{1}{2}\sum_{i, j}^{N}J_{ij}s_{i}s_{j}+\sum_{i=1}^{N}h_{i}s_{i}\right)\equiv\frac {e^{ \mathcal{U}(\textbf{s})}}{\mathcal{Z}}\label{Lagrange}
\end{equation}

where $J_{ij}$ and $h_{i}$ are Lagrange multipliers (chosen to retrieve the first and second empirical moments). They can be thought as a measure of the pairwise mutual and individual influences.
Another well known application of the pairwise maxent model is the characterization of the neural network structure \cite{Schn} where the operating point seems to be a critical one \cite{Mora,fraiman}. One can show that this model is able to generate correlation matrices with non-Gaussian eigenvalues \cite{moi1} as observed in real financial time series \cite{refLaloux} but also scale-free asset trees and order-disorder periods \cite{moi2}. This pairwise model gives more insights about the possibility of a critical operating point.
The rescaling of the Gibbs distribution is then viewed as a rescaling of all the parameters by a common factor $T^{-1}$. This rescaling is an investigation of a slice of the parameters space which corresponds to a stochasticity variation. A small value of $T$ favoris co-movements and a large value favoris the randomness. In this work, Lagrange multipliers are estimated with a regularized pseudo-maximum likelihood \cite{Aurell}, see \ref{sec:rPML} for a short description. We note that close to $T=1$, many models are distinguishable and a slight change in parameters may lead to a significant change of the measured observables. One should compare artificial and empirical results.

%

\begin{figure}[!ht]
\begin{center}
\includegraphics[width=\textwidth]{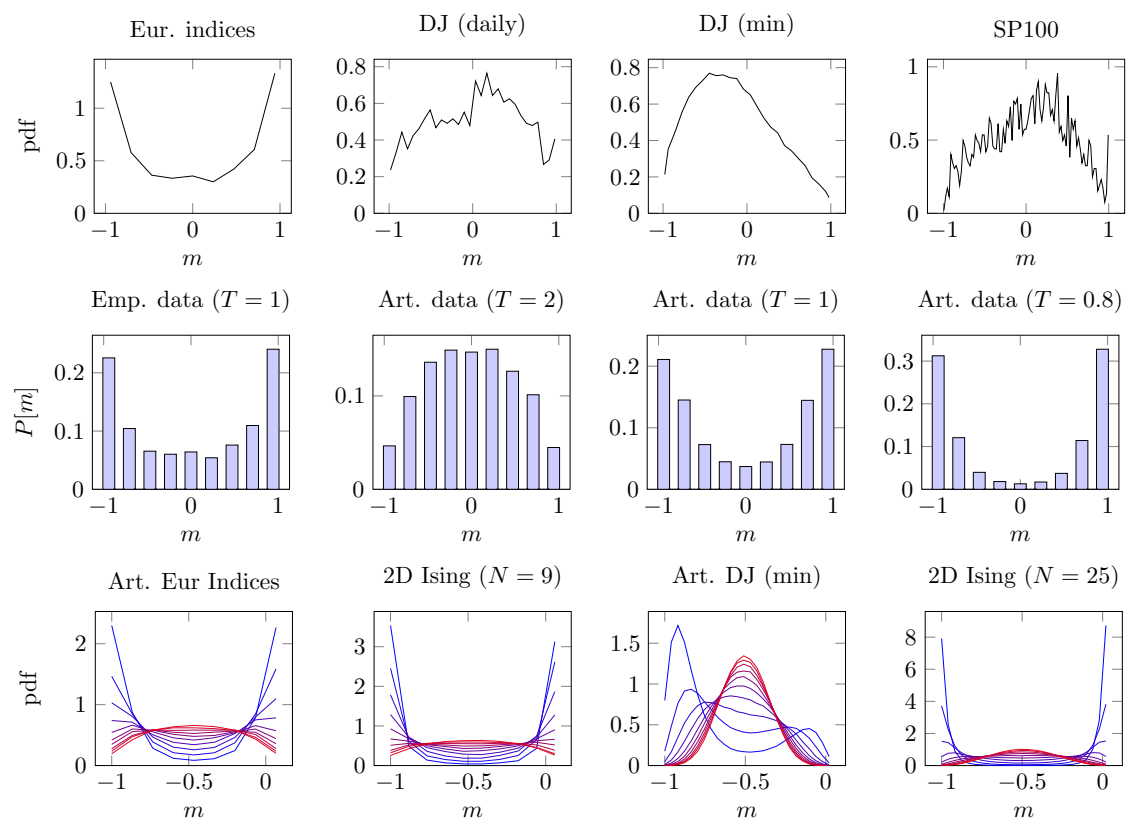}
\end{center}
\caption{First row: the empirical probability density function (pdf) is illustrated for several data sets. Second row: comparison of the empirical probability mass function (pmf) of the net orientation to the artificial distributions resulting from simulations. Third row: 10 values of the stochasticity level $T$ (in the range $[0.8,2]$, blue to red respectively) are used to check if the pdf can go continuously from unimodal to bimodal, the results are compared to a 2D nearest neighbour Ising model without individual biases. The pdf and the pmf are estimated on $5\times 10^{5}$ Monte Carlo steps.}
\label{fig:Pm}
\end{figure}

First, we simulate artificial data with the estimated Lagrange parameters from the real time series. The Monte Carlo Markov chain (MCMC) is defined as follows. A randomly chosen orientation is flipped if the conditional flipping probability $p(s_{i,t}=-s_{i,t-1}|s_{-i,t})$ is larger than a realization of a uniform law on the interval $[0,1]$, where $s_{-i,t}$ is the configuration excluding the $i$th entity. A configuration is recorded each $N$ flipping attempts, which defines a Monte Carlo step (MCS) \cite{Glauber}. See \ref{sec:MCMC} for details.
The result of the procedure applied to those artificial data is illustrated in Fig-\ref{fig:hc} by the dashed curve ($1\times 10^4$ equilibration MCS and $1\times10^5$ recorded MCS). This is consistent with the empirical variance, both peaks (blue and dashed curves) are located at the same value of the $T$-parameter.
If Lagrange parameters $\{J_{ij}\}$ are positive, the orientation distribution should be unimodal for large value of $T$ and bimodal for small value of $T$. As a qualitative test, we check if the empirical distributions are unimodal or bimodal and if they can become bimodal if we change the stochasticity level $T$, an order-disorder transition is then possible. As illustrated in the first row of Fig-\ref{fig:Pm}, the empirical distribution of the indices set is bimodal whereas the distributions of stock sets are unimodal as expected from the former empirical analyses. The second row of Fig-\ref{fig:Pm} illustrates the difference between the empirical orientation distribution and the simulated ones $P_{m}(T)$ at different stochasticity levels. The indices set is a rather ordered system, the probability mass peaks at the extremes values $-1,1$ of the net orientation. A disordered state exists for high level of stochasticity ($T=2$). The third row of Fig-\ref{fig:Pm}  illustrates the continuous deformation of the probability density function for a stochasticity varying from low level (blue) to high level (red). This deformation is compared to the one of the 2D nearest neighbor Ising model of corresponding size without individual biases.

\begin{figure}[ht!]
\begin{center}
\includegraphics[width=\textwidth]{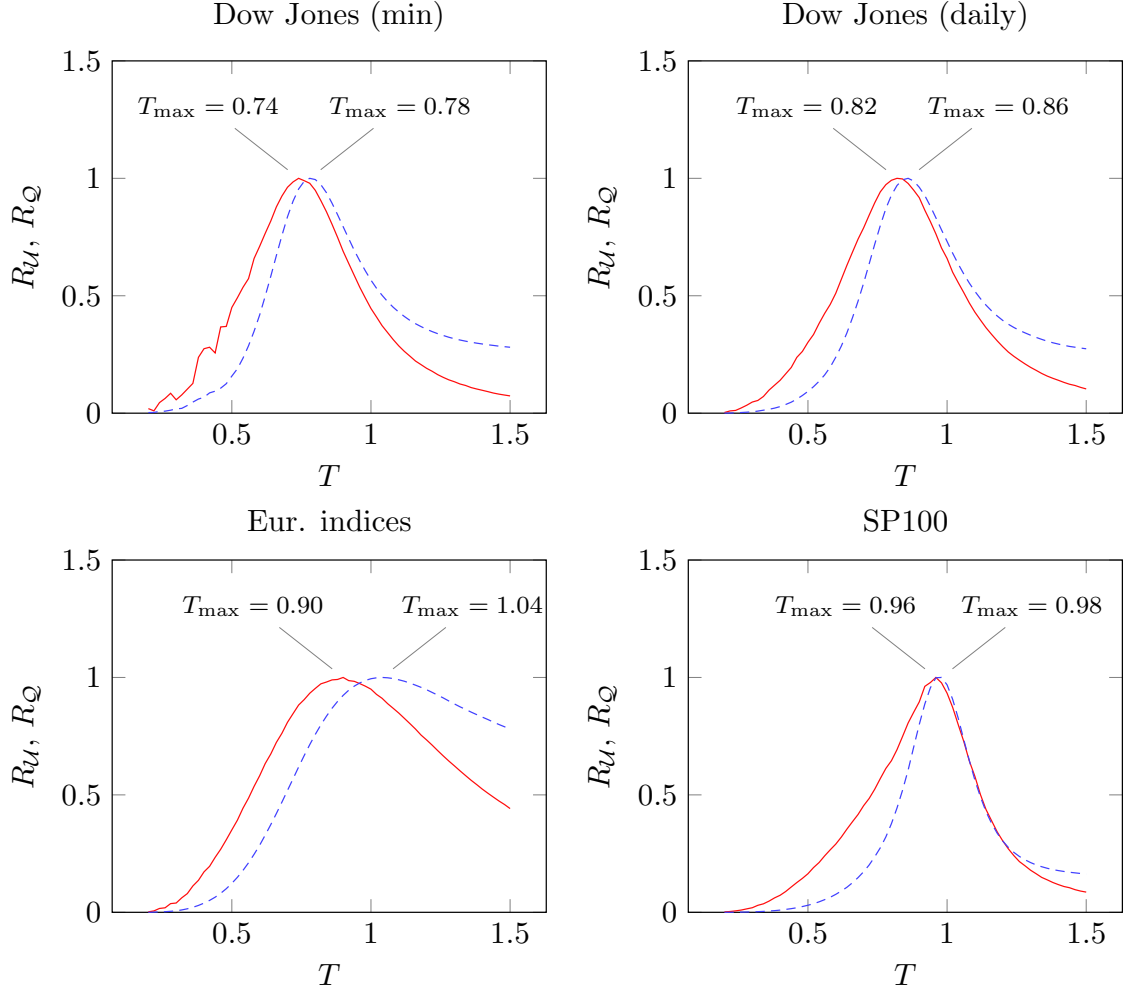}
\end{center}
\caption{Variances of the overlap parameter (dashed lines) and of the log-likelihood (full lines) for the 8 indices set, Dow Jones (daily and minute samplings) and SP100. Each point is computed over $5\times 10^5$ MCS after an equilibration period of $5\times 10^4$ MCS. The coordinate of the maximum is pinned (the coordinate on the left stands for the variance of the log-likelihood) .}
\label{fig:chi}
\end{figure}

The fitted maxent models allow an order-disorder transition which justifies their use in the criticality check. As mentioned in \cite{Mastro} such models are prone to accumulate in the vicinity of the critical point $T=1$ but are also highly distinguishable in this neighbourhood. Accordingly, we check if they return a $T_{\mathrm{max}}$ in line with the empirical results. One can estimate the variance $R_{\mathcal{Q}}$ of the overlap parameter $q=N^{-1}\sum_{i}s_{i}^{(1)}s_{i}^{(2)}$ and the variance of the log-likelihood. The overlap parameter measures the correlation between the configurations of two identical systems denoted by the superscript (1) and (2). The variances $R_{\mathcal{U}}$ and $R_{\mathcal{Q}}$ are known to peak at the critical value of the rescaling parameter \cite{Fisher}. If the operating point is indeed critical, we should find the peak near the value $T=1$. The results are illustrated in Fig-\ref{fig:chi}.

We note that the peaks are indeed located near the empirical values. For the indices set, the relative difference between empirical and simulated $T_{\mathrm{max}}$ is equal to $2\%$, slightly underestimated. For the Dow Jones (min), the relative difference is equal to $6\%$, slightly overestimated and for the Dow Jones (daily), $T_{\mathrm{max}}$ is overestimated of $14\%$.
The first two fitted models are consistent with the data and lead to the same conclusion: the indices set is close to the criticality ($1-T_{\mathrm{max}}\leq 10\%$) and the Dow Jones is far from criticality ($1-T_{\mathrm{max}}\geq 25\%$).
The larger deviation between empirical and simulated values for the Dow Jones (daily) may be due to inference errors in the Lagrange parameters estimation. The ratio $M/N$ (sample length on the number of entities) is too small, ten times smaller than for the Dow Jones (min). Consequently, one may expect the same relative error for the critical scaling parameter of the SP100 index.

\begin{figure}[ht!]
\begin{center}
\includegraphics[scale=1]{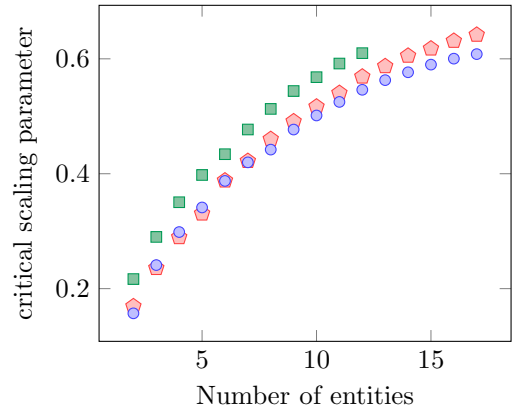}
\end{center}
\caption{Value of $T$-parameter at which response function $R_{\mathcal{U}}$ reaches its maximum value vs the number of entities. The squares illustrate the real data, the pentagons stand for a multivariate $\mathrm{GARCH}(2,2)$ process and the dots for the MCMC.}
\label{fig:TcriArt}
\end{figure}

Since simulations are consistent with empirical results, we simulate data to complete Fig-\ref{fig:Tcri} for sizes larger than $N=8$. We simulate a binary sample of length $5\times10^6$ with the previous MCMC and also artificial returns with a multivariate $\mathrm{GARCH}$ process, known to capture clustering volatility and fatter tail than Gaussian one. We obtain results consistent with the empirical ones. The critical value of the rescaling parameter $T_{\mathrm{crit}}$ is illustrated in Fig-\ref{fig:TcriArt}. The critical value increases with size but is still far from $T=1$.

We note that in 2010, 12807 companies (excluding investment funds) have been listed in stock exchanges (see \url{http://www.world-exchanges.org/}). There is thus no obvious reason to consider the limit $N\rightarrow\infty$.
The market places system is significantly closer to the criticality despite its small size. It may be due to information aggregation of an index about the underlying stocks \cite{Shap}. A set of indices may operate as a system of larger size.

Furthermore, one can show that the financial network exhibits small-world organization \cite{Petra} and one knows that the Ising model on a complex network, among other, is a small-world one only at the critical temperature \cite{fraiman}.



\section{Discussion}

Stock markets are embedded in a non-uniform background. They should therefore be heterogeneous and go through regular periods interspersed with surprising events.  In a complex economic background, reactiveness is an expected behaviour. In the case of the Fukushima nuclear accident or the 2008 subprime crisis for instance, the market response was clear and prompt. All stocks fell quickly in an organized fashion. This behaviour can help to secure the profit made or prevent excessive losses if the situation goes even worse. Then, when the situation seems stabilized, or that stocks prices have fallen so dramatically that stocks became cheap and attractive, the market goes up again in an ordered fashion. These large bearish-bullish movements of the stock prices are encountered at any time scale \cite{Preis}. During such phases, the market exhibits large correlated structures and ordered state \cite{Dal,Jr,moi2} corresponding to an increase of the correlation strength. Such dramatic events impact globally the market (all economic sectors). On the other hand, some events (like the end of a state subsidy for eco-friendly goods, nuclear energy, etc.) have an impact on a single or few economic sectors. The criticality is then thought as a competition between global effects inducing homogeneity and local effects inducing heterogeneity in trades.

We have seen that Shannon entropy has an inflexion point near the operating point $T=1$ for the European indices set. We deduce that the micro-states number increases (or decreases) drastically following a variation of the stochasticity. The entropy is related to the logarithm of the averaged micro-states number and we can obtain this quantity by a simple integration of $R_{\mathcal{U}}(T)/T$. We observe that the largest slope stands approximatively at the actual operating point $T=1$ far from the saturation zones (where the slope is close to zero). In the neighborhood of the operating point, the logarithm of the number of micro-states is almost linear with a large slope thus a variation of Lagrange parameters will induce a drastic (in an exponential fashion) change in the micro-structure.  It shows that the market network has a great structural malleability. The entropy also measures the degree of statistical dependency between stocks. If stocks did not influence each other, the system would be considered as a random one which implies small covariances and low reactiveness. Thus entropy would reach its largest value. In the opposite case, if stocks correlations are maximal (implying again low reactiveness), there would not be any incertitude anymore, the whole market state $\mathbf{s}$ would be predictable on the knowing of a individual state $s_{i}$ and the entropy would be zero. So if the slope of the entropy reaches its maximum value at the operating point, it means that the market is on the edge. Any variation can tip the market either towards a random (disordered, with independent trades) either towards a highly interactive (ordered, synchronized trades) state. We expect thus a large predictability exploiting instantaneous information: using the system configuration amputated of the $i$th entity $\mathbf{s}_{-i}$, one should be able to predict the state of this entity $s_{i}$ with high accuracy. This will be the subject of another work.

\begin{figure}[ht!]
\begin{center}
\includegraphics[width=\textwidth]{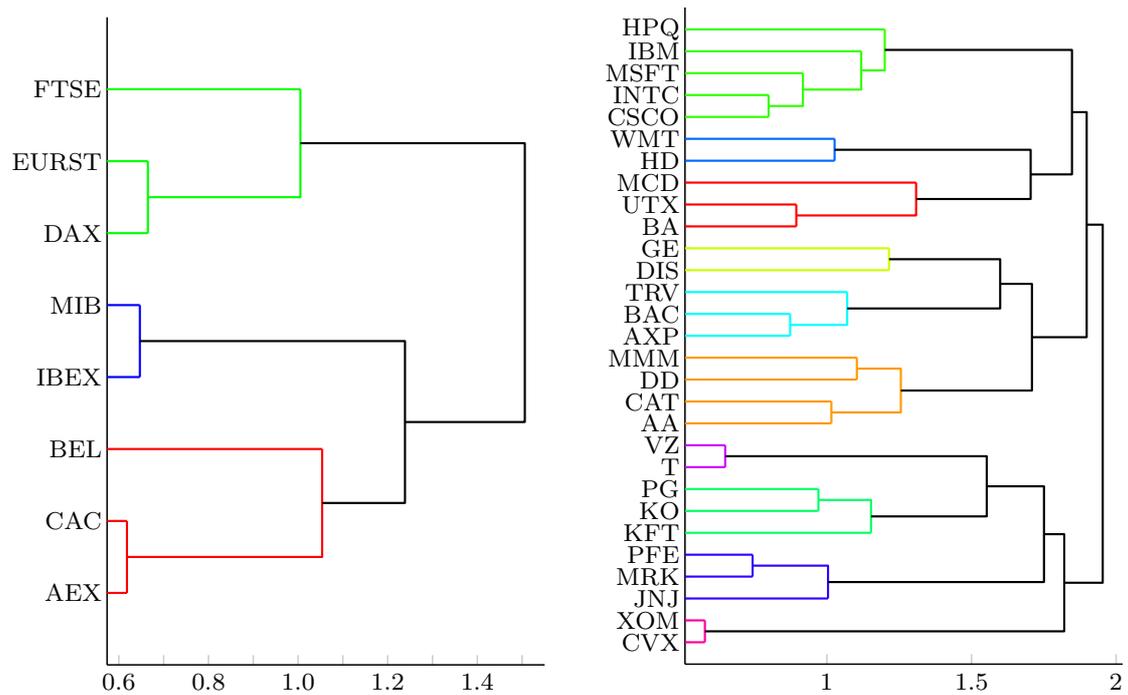}
\end{center}
\caption{Illustration of the clusters of each data sets. The clusters of the indices set (left) returns a partition of the European economy. The clustering of the Dow Jones (right) returns the different economic sectors (technologies, distribution, aircraft industry, TV broadcasting, finance, chemical and industrial companies, telecom, consumer goods, health care, oil).}
\label{fig:Clusters}
\end{figure}

Last, the fact that the European indices set is closer to the criticality than the Dow Jones may follow from information aggregation \cite{Shap}. A set of indices is a weighted average of stock prices. Considering the stocks as the fundamental hubs of the financial network, the indices represent super-hubs acting as a system of significantly larger size. The typical relative cluster size is also larger in the indices set where each cluster contains roughly $30\%$ of the total number of entities as illustrated in Fig-\ref{fig:Clusters} \cite{moi2}. For the Dow Jones, the cluster size is about $10\%$ of the index size. Correlated structures have thus a larger relative size in the indices set which may match the right balance between co-movements and fluctuations.

From the data analysis and simulations, we saw that the European market places seem to operate at a point where the variances of the log-likelihood is close to their largest values. An exponential empirical fit returns $T_{\mathrm{op}}=0.92$ as asymptotical value (thus maximum) for European indices and $T_{\mathrm{op}}=0.70$ for the Dow Jones at minute sampling. The entropy is not a linear function of the log-likelihood. The estimation of $T_{\mathrm{op}}$ with simulated data returns a value close to one but this value is suspected to be overestimated about $15\%$. For the Dow Jones, large simulated samples $M=5\times 10^{6}$ (using parameters obtained by fitting real data) return a consistent value $T_{\mathrm{op}}\simeq 0.65$.
Moreover, financial systems are closer to the criticality close to the crash beginning meaning large fluctuation and large deviation from the uniform distribution of the configurations. This evolution also suggests a process of self-organization. The market is a highly adaptive system. By self-organization, the market reacts strongly to a change or unexpected events and by itself does not consider all possible events as equiprobable. However through the data analysis, the stock exchanges system is not exactly critical and the Dow Jones seems to be far from criticality. Furthermore, financial systems do not stay in the same regime and get closer to the criticality just before a crisis. An interesting finding since in such models, large avalanches occur more likely close to the criticality \cite{Perkovic}.

\section*{Acknowledgments}
I would like thank B. De Rock and P. Emplit for their helpful comments and discussions. This work was undertaken with financial support from the Solvay Brussels School of Economics and Management.

\appendix

\section{Practical recipe}

\begin{enumerate}
\item Binarize the returns.
\item Test the statistical significance and determine the corresponding maximum size $N$.
      \begin{enumerate}
         \item Compute the empirical distribution of configurations $\hat{p}_{\mathbf{s}}$.
         \item Compute $m_{k}$ (the number of configurations sampled exactly $k$ times) and the empirical distribution of $K_{i}$ (the number of times the configuration $\mathbf{s}_{i}$ is observed in the sample).
         \item Deduce their entropies $H[\mathbf{s}]$ and $H[K]$.
         \item Locate the maximum of the relation  $H[\mathbf{s}]$ vs $H[K]$.
      \end{enumerate}
\item Get the response function $R_{\mathcal{U}}$ and find its maximum. Repeat for several ($\sim 100$) sets of $N$ randomly chosen entities. For each set:
      \begin{enumerate}
         \item Compute the empirical distribution of configurations $P(\mathbf{s})$.
         \item Rescale the empirical distribution as $P_{T}(\mathbf{s})=\frac{P(\mathbf{s})^{1/T}}{\sum_{\{\mathbf{s}\}}P(\mathbf{s})^{1/T}}$.
         \item Compute $R_{\mathcal{U}}=T \frac{\partial \mathcal{S}}{\partial T}$ where $\mathcal{S}(T)=-\sum_{\{\mathbf{s}\}}P_{T}(\mathbf{s})\log P_{T}(\mathbf{s})$.
         \item Store the coordinates of its maximum.
      \end{enumerate}
\item Compare to a finite size version of a truly critical system.
       \begin{enumerate}
         \item Compute the relative difference $x=(T_{\mathrm{op}}-T_{\mathrm{max}})/T_{\mathrm{op}}$ where $T_{\mathrm{op}}=1$.
         \item Compute the Kullback-Leibler divergence (KLD) between $P_{T=T_{\mathrm{max}}}(\mathbf{s})$ and $P_{\mathrm{emp}}(\mathbf{s})$.
         \item Compare the latter KLD value to the KLD between $P_{T=T_{\mathrm{crit}}}(\mathbf{s})$ and $P_{T=(1+x)T_{\mathrm{crit}}}(\mathbf{s})$ for the 2D nearest neighbours Ising model.
       \end{enumerate}
\item Perform a statistical test of Zipf's law as described in \ref{sec:DPL}.
\item Check the linearity of the relation $\mathcal{S}(\mathcal{U})$ vs $\mathcal{U}$ where $\mathcal{U}=\log P(\mathbf{s})$.
\item Compare the empirical results to simulations.

   \begin{enumerate}
         \item Infer the Lagrange parameters (see \ref{sec:rPML}).
         \item Simulate data using a Monte Carlo Markov chain (see \ref{sec:MCMC}).
         \item Check if an order-disorder transition is allowed by computing the orientation distribution and by varying the scaling parameter $T$.
         \item Compute the variance of the log-likelihood and of the overlap parameter $q=N^{-1}\sum_{i}s_{i}^{(1)}s_{i}^{(2)}$ (two copies denoted by the superscript, linked with the covariance of the utility function $\mathcal{U}$). Compare empirical and simulated results for a common size. A large difference ($>10\%$) between the simulated value of $T_{\mathrm{max}}$ and the asymptotical value returned by fitting the empirical relation $T_{\mathrm{max}}(N)$ may reveal  difficulties in the inference of Lagrange parameters \cite{Mastro} and therefore a poor fitting.
       \end{enumerate}
\end{enumerate}

\section{Discrete power-law}\label{sec:DPL}
A statistical test for power-law is given in \cite{Clauset}. We adapt this test to discrete power-law with a natural upper bound. Before considering the discrete case, we note that if the distribution $p(x)\sim x^{-\beta}$ has a finite upper bound $x_{\mathrm{max}}$, then the cumulative distribution function (CDF) will not be a straight line in a log-log plot because

\begin{equation}\label{3-PLCDF}
  P[X\geq x]=\mathrm{Cst} \int_{x}^{x_{\mathrm{max}}}y^{-\beta}\ud y= \frac{\mathrm{Cst}}{1-\beta}
  \left[x_{\mathrm{max}}^{1-\beta}-x^{1-\beta} \right]
\end{equation}
where the constant normalizes the distribution to 1 and $\beta>1$. Taking the logarithm of both sides, it comes

\begin{equation}
  \log P[X\geq x]= \log\left(x^{1-\beta}-x_{\mathrm{max}}^{1-\beta}\right)+\log \frac{\mathrm{Cst}}{\beta-1}
\end{equation}

The statistical test proposed in \cite{Clauset} consists to the following scheme

\begin{enumerate}
  \item Determine the best fit of the power-law to the data using maximum-likelihood estimator.
  \item Calculate the Kolmogorov-Smirnov (KS) statistics for the goodness-of-fit. The KS statistics is the maximum absolute value between empirical CDF and the CDF of the estimated power-law.
  \item Generate a large number ($\sim 1000)$ of synthetic data sets.
  \item Calculate the p-value as the fraction of the KS statistics for the synthetic data sets whose value exceeds the KS statistics of the real data.
  \item If the p-value is sufficiently small ($\sim 0.05$), the power-law is ruled out.
\end{enumerate}

The MLE estimator of a discrete power-law with a natural cut-off $x_{\mathrm{max}}$ is derived from the first order condition for the log-likelihood based on $N$ observations

\begin{equation}
  \mathcal{L}(\beta)=\ln L(\beta)= -\beta \sum_{i=1}^{N}\ln x_{i}-N \ln\left(\sum_{x=1}^{x_{\mathrm{max}}}x^{-\beta}\right)
\end{equation}
taking the derivative with respect to $\beta$ leads to the MLE $\beta_{\mathrm{MLE}}$ satisfying

\begin{equation}\label{PLMLE}
  \frac{1}{N}\sum_{i=1}^{N}\ln x_{i}=  \frac{\sum_{x=1}^{x_{\mathrm{max}}}x^{-\beta_{\mathrm{MLE}}}\ln x_{\mathrm{max}}}{\sum_{x=1}^{x_{\mathrm{max}}}x^{-\beta_{\mathrm{MLE}}}}
\end{equation}

The standard deviation of $\beta_{\mathrm{MLE}}$ is obtained by taking the expansion of the likelihood around $\beta_{\mathrm{MLE}}$

\begin{equation}
  \mathcal{L}(\beta) = \mathcal{L}(\beta_{\mathrm{MLE}})+\frac{1}{2!}\frac{\partial^{2}\mathcal{L}(\beta)}{\partial\beta^{2}}\Big|_{\beta_{\mathrm{MLE}}}
  (\beta-\beta_{\mathrm{MLE}})^{2}
\end{equation}
identifying the terms to the Gaussian approximation $-\ln(\sigma \sqrt{2\pi})-\frac{1}{2}\left(\frac{x-\beta}{\sigma}\right)^{2}$, it comes

\begin{equation}
  \sigma_{\beta_{\mathrm{MLE}}}=\frac{1}{\sqrt{N\left[\frac{\zeta^{''}(x_{\mathrm{max}},\beta_{\mathrm{MLE}})}
  {\zeta(x_{\mathrm{max}},\beta_{\mathrm{MLE}})}-\left(\frac{\zeta^{'}(x_{\mathrm{max}},\beta_{\mathrm{MLE}})}
  {\zeta(x_{\mathrm{max}},\beta_{\mathrm{MLE}})}\right)^{2}\right]}}
\end{equation}
where $\zeta(x_{\mathrm{max}},\beta)=\sum_{x=1}^{x_{\mathrm{max}}}x^{-\beta}$ and the prime stands for the derivative with respect to $\beta$.

Synthetic data distributed as a discrete power-law with a finite upper bound are generated as follows. One generates a realization $u$ of a uniform random variable $U$ in $[0,1]$, one calculates $\sum_{x=1}^{x_{\mathrm{max}}}x^{-\beta}$ and the cumulative sum $\sum_{y=1}^{x}y^{-\beta}$. The smallest integer $x$ such that $\sum_{y=1}^{x}y^{-\beta}\geq u \, \sum_{x=1}^{x_{\mathrm{max}}}x^{-\beta}$ is stored. This process is repeated to generate a sample of desired length.

\section{Regularized pseudo-maximum likelihood}\label{sec:rPML}

The rPML method is a powerful method for estimation of Lagrange parameters of pairwise maximum entropy model while common maximum likelihood is untractable \cite{Aurell}. This method can be thought as an autologistic regression in order to predict binary outcomes. The main idea is to factorize the distribution and to consider only conditional probabilities. For a N-dimensional sample of length $M$, the objective function to maximize is

\begin{equation}\label{PML}
  \mathrm{PL}(\boldsymbol\theta)=\frac{1}{M}\sum_{t=1}^{M}\sum_{i=1}^{N}
  \log P(s_{i,t}|\mathbf{s}_{-i,t};\, \boldsymbol\theta)
\end{equation}
where $\mathbf{s}_{-i}$ stands for the configuration excluding the $i$th entity and the conditional probabilities of the pairwise maximum entropy model are

\begin{equation}
p(s_{i,t}|\mathbf{s}_{-i,t};\, \boldsymbol\theta)=
\frac{1}{2}
\left[1+s_{i,t}\tanh\left(\sum_{ j\neq i}J_{ij}s_{j,t}+h_{i}\right)\right]
\end{equation}

A regularization term is added to the PL function to prevent overfitting which is a negative multiple of the $l_{2}$-norm of parameters to be estimated, for instance. The regularized PL (rPL) objective function is thus $\mathrm{PL}(\boldsymbol\theta)-\lambda\, \|\boldsymbol\theta\|_{2}$ with $\lambda>0$.

\section{Monte Carlo Markov chain simulations}\label{sec:MCMC}

To perform simulation, we should describe how the Gibbs distribution can be reached as the equilibrium distribution of a given Markov process. A way to reach a Gibbs distribution

\begin{equation}
p_{2}(\textbf{s})=\mathcal{Z}^{-1}\exp\left(\frac{1}{2}\sum_{i, j}^{N}J_{ij}s_{i}s_{j}+\sum_{i=1}^{N}h_{i}s_{i}\right)\equiv\frac {e^{- \mathcal{H}(\textbf{s})}}{\mathcal{Z}}\label{3-Gibbs}
\end{equation}
is given by the following dynamics (the so-called Glauber dynamics \cite{Glauber}). Namely, one takes an entity $i$ chosen randomly and the attempt to flip the associated binary variable $s_{i}$ is performed with a rate depending on an exponential weight, the other orientations remaining fixed. We define the reversal operator $\Flip{i}$ such that
$\Flip{i}\textbf{s}=\Flip{i}(s_{1},\ldots,s_{i},\ldots,s_{N})=(s_{1},\ldots,-s_{i},\ldots,s_{N})$. This asynchronous updating involves that two consecutive configurations only differ by a single reversal.
To find the exponential rate, we consider the evolution of the probability mass function (PMF) for this dynamics which is given by the master equation

\begin{equation}
\frac{\mathrm{d}}{\mathrm{d}t}p(\textbf{s}; t)=
\sum_{i = 1}^{N} \Big\{\omega(s_{i}|\, -s_{i})\;p(\Flip{i}\textbf{s}; t)-\omega(-s_{i}|\, s_{i})\;p(\textbf{s}; t)\Big\}\label{3-Master}
\end{equation}
where $\omega(s_{i}|\, -s_{i})$ is the transition rate from configuration $\Flip{i}\textbf{s}$ to configuration $\textbf{s}$.
They are derived from the transition probability $\mathrm{P}[s_{i,t+\tau}=-s_{i,t}|s_{i,t},\, \mathbf{s}_{-i,t}] \equiv W(-s_{i}|s_{i},0) = \omega(-s_{i}|\, s_{i})\,\tau+o(\tau)$.

The master equation states that the variation of the PMF is equal to the inward probability flow minus the outgoing probability flow \cite{Kampen}. At equilibrium, this dynamics should lead to the Gibbs distribution (\ref{3-Gibbs}). A sufficient condition to reach equilibrium is

\begin{equation}\label{3-DBal}
  \omega(s_{i}|\, -s_{i})\;p_{2}(\Flip{i}\textbf{s})-\omega(-s_{i},|\, s_{i})\;p_{2}(\textbf{s})=0
\end{equation}
As we are only interested on the equilibrium PMF and not how one reaches it, we can choose any transition rates satisfying (\ref{3-DBal}). A convenient choice for simulation (discrete time) is to take the transition probability

\begin{equation}\label{3-TP}
   W(-s_{i}|s_{i})= \frac{1}{2}\left[1-s_{i,t}\tanh\left(\sum_{j}J_{ij}s_{j,t}+h_{i}\right)\right]
\end{equation}

Simulations are performed following the scheme\newline

\textbf{Algorithm}
\begin{enumerate}
  \item Choose an entity uniformly at random.
  \item Compute the transition probability (\ref{3-TP}).
  \item Generate a uniform random number $x\in[0,1]$, if $W(-s_{i}|s_{i})>x$, accept the reversal.
  \item Parameterize time such that a Monte Carlo step (MCS) corresponds to $N$ reversal attempts.
  \item Wait for equilibration.
  \item Store the desired statistics.
\end{enumerate}

A more detailed discussion (equilibration time, proper definition of statistics, etc.) can be found in \cite{BinderMC}.


%

\end{document}